\documentclass[amsmath,amssymb,prl,hyperlink,twocolumn,longbibliography]{revtex4}
\usepackage{graphicx}
\usepackage{soul}
\usepackage[colorlinks=true,citecolor=blue,linkcolor=magenta]{hyperref}
\usepackage[usenames]{color}
\usepackage{amsfonts}
\usepackage{color}
\usepackage{booktabs}
\usepackage{multirow}
\usepackage{float}
\usepackage[british]{babel}
\usepackage[style=american]{csquotes}
\usepackage{times}
\usepackage{epstopdf}
\usepackage{lineno}
\usepackage{float}
\usepackage{ulem}

\newcommand{\ketbra}[1]{\ensuremath{| #1 \rangle \langle #1 |}}
\newcommand{\ket}[1]{\ensuremath{|#1\rangle}}

\newcommand{\trace}{{\rm Tr}}

\newcommand{\blue}[1]{{\color{black} #1}}

\begin{document}

	\title{{Verification of a resetting protocol for an uncontrolled superconducting qubit}}
	\author{Ming Gong,$^{1,2,3}$ Feihu Xu,$^{1,2,3}$ Zheng-Da Li,$^{1,2,3}$ Zizhu Wang,$^{4}$ Yu-Zhe Zhang,$^{1,2,3}$ Yulin Wu,$^{1,2,3}$ Shaowei Li,$^{1,2,3}$  Youwei Zhao,$^{1,2,3}$ Shiyu Wang,$^{1,2,3}$ Chen Zha,$^{1,2,3}$ Hui Deng,$^{1,2,3}$ Zhiguang Yan,$^{1,2,3}$ Hao Rong,$^{1,2,3}$  Futian Liang,$^{1,2,3}$ Jin Lin,$^{1,2,3}$ Yu Xu,$^{1,2,3}$ Cheng Guo,$^{1,2,3}$ Lihua Sun,$^{1,2,3}$ Anthony D. Castellano,$^{1,2,3}$ Cheng-Zhi Peng,$^{1,2,3}$  Yu-Ao Chen,$^{1,2,3,}$ Xiaobo Zhu,$^{1,2,3,}$\footnote{Corresponding author. Email: xbzhu16@ustc.edu.cn} and Jian-Wei Pan$^{1,2,3}$\vspace{0.2cm}}
	\affiliation{$^1$  Hefei National Laboratory for Physical Sciences at the Microscale and Department of Modern Physics, University of Science and Technology of China, Hefei 230026, China}
	\affiliation{$^2$  Shanghai Branch, CAS Center for Excellence in Quantum Information and Quantum Physics, University of Science and Technology of China, Shanghai 201315, China}
	\affiliation{$^3$  Shanghai Research Center for Quantum Sciences, Shanghai 201315, China}
	\affiliation{$^4$  Institute of Fundamental and Frontier Sciences, University of Electronic Science and Technology of China, Chengdu 610054, China}
	
	%
	
\begin{abstract}	
Quantum resetting protocols allow a quantum system to be sent to a state in the past by making it interact with quantum probes when neither the free evolution of the system nor the interaction is controlled. We experimentally verify the simplest non-trivial case of a quantum resetting protocol, known as the $\mathcal{W}_4$ protocol, with five superconducting qubits, testing it with different types of free evolutions and target-probe interactions. After projection, we obtained a reset state fidelity as high as $0.951$, and the process fidelity was found to be $0.792$. We also implemented 100 randomly-chosen interactions and demonstrated an average success probability of $0.323$ for $|1\rangle$ and $0.292$ for $|-\rangle$, experimentally confirmed the nonzero probability of success for unknown interactions; the numerical simulated values are about $0.3$. Our experiment shows that the simplest quantum resetting protocol can be implemented with current technologies, making such protocols a valuable tool in the eternal fight against unwanted evolution in quantum systems.
\end{abstract}

\maketitle
\begin{figure*}[ht]
	\centering
		\includegraphics[width=0.7\textwidth]{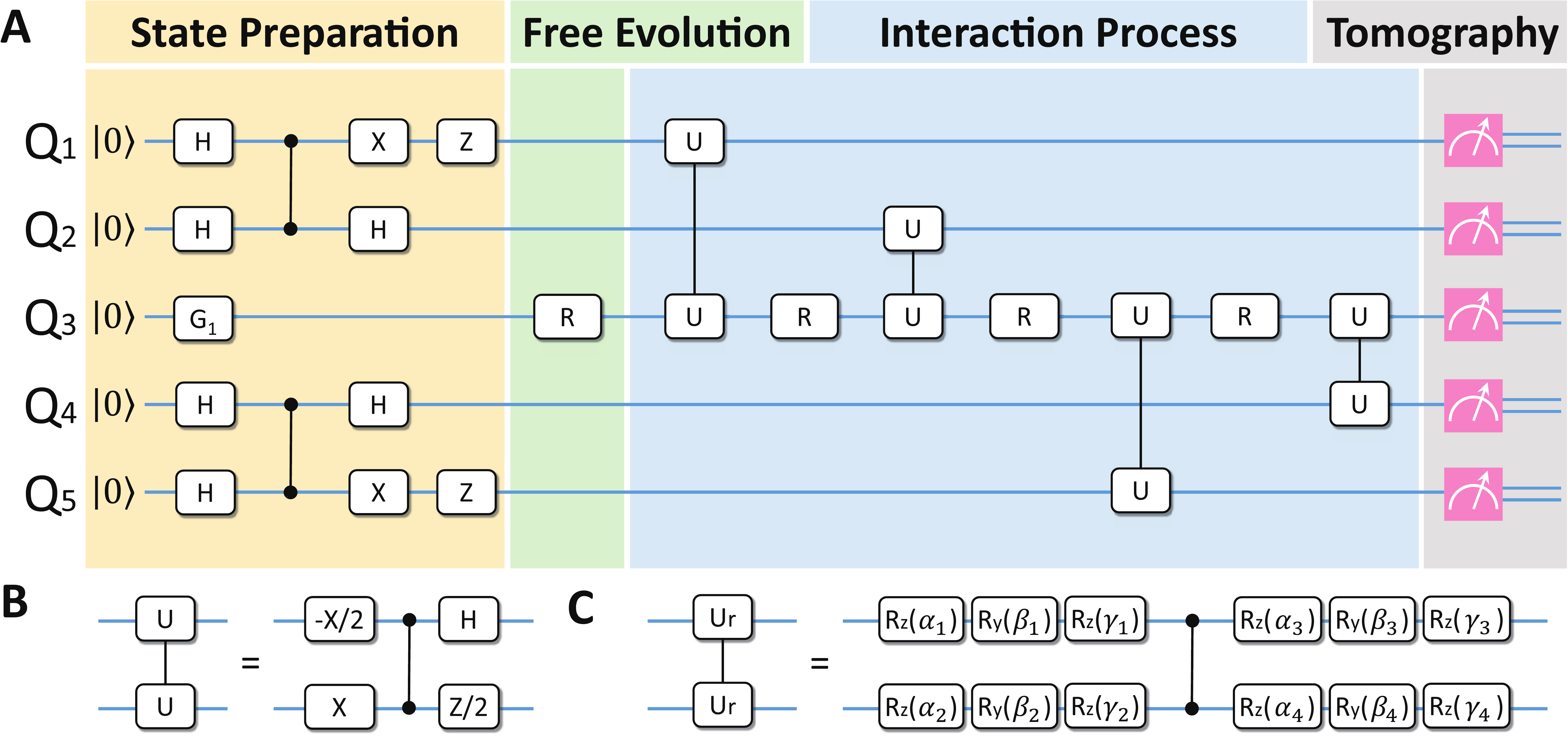}
	\caption{\textbf{Description of the quantum resetting protocol.} 
		\textbf{A.} The quantum circuit. During state preparation (yellow), the target qubit $Q_3$ is set with the single-qubit gate $G_1$, and the probe pairs $Q_{1,2}$ and $Q_{4,5}$ are prepared in the Bell state $|\psi^-\rangle$. The single qubit gate $R$ is then applied on the target qubit during the evolution phase (green). The probes are then made to interact with the target, with the bipartite unitary operator $U$ governing the interaction. We note that the bipartite operation between the next-nearest-neighboring qubits $Q_1$ and $Q_3$ is realized as the bipartite operation between $Q_2$ and $Q_3$ and then the SWAP operation between $Q_1$ and $Q_2$ (see Method for details), so as the operation between $Q_5$ and $Q_3$. Finally, the five-qubit combined state is obtained via quantum state tomography, which is then projected for a successful reset. \textbf{B.} {Realization of a} gate sequence for the deterministic interaction operator $U=(X\otimes Z+iY\otimes X)/\sqrt{2}$. \textbf{C.} {Realization of} random unitary interaction. Three single-qubit rotation gates perform an arbitrary rotation, the rotation angles of which are chosen randomly. Each rotation is applied on target and probe before and after the $CZ$ gate.}
	\label{circuit}
\end{figure*}

\begin{figure*}[htb]
	\centering
		\includegraphics[width=0.6\textwidth]{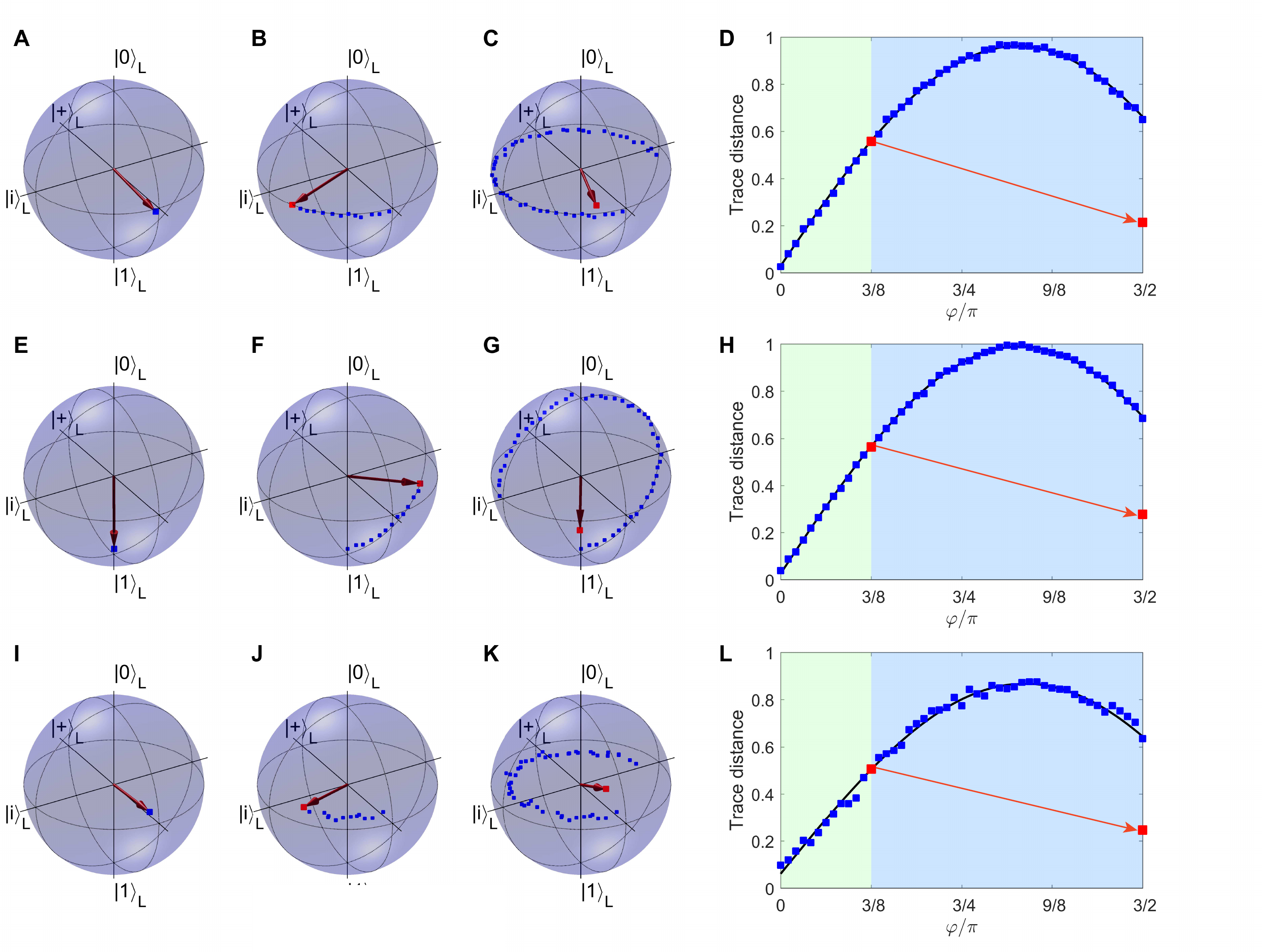}
	\caption{\textbf{Resetting the target qubit after a free evolution.} 
		On the Bloch spheres, the evolution of the states for the three phases are presented: After state preparation (A, E, I), after free evolution (B, F, J), and after resetting (C, G, K). The resetting protocol is applied when the free-evolution induced phase is $\varphi=3\pi/8$. The red dots mark the state before and after a successful reset. The blue dots mark a free evolution without resetting as a comparison, simulating the results of the time-independent Hamiltonian $\mathcal{H}_0$. As the resetting protocol requires three more free evolutions, the state without the resetting process is measured till $\varphi=3\pi/2$. From top to bottom, each row shows a different version of the resetting protocol {in case 1, demonstrated the resetting process for a superposition state \ket{-}, a classical state \ket{1}, and a mixed state, respectively}.
		In D, H, and L, {the trace distance after the application of resetting protocol are observed jumping from 0.557, 0.564, and 0.506, to 0.214, 0.277, and 0.246, respectively.}}
	\label{proof}
\end{figure*}

\begin{figure*}[ht]
	\centering
		\includegraphics[width=0.6\textwidth]{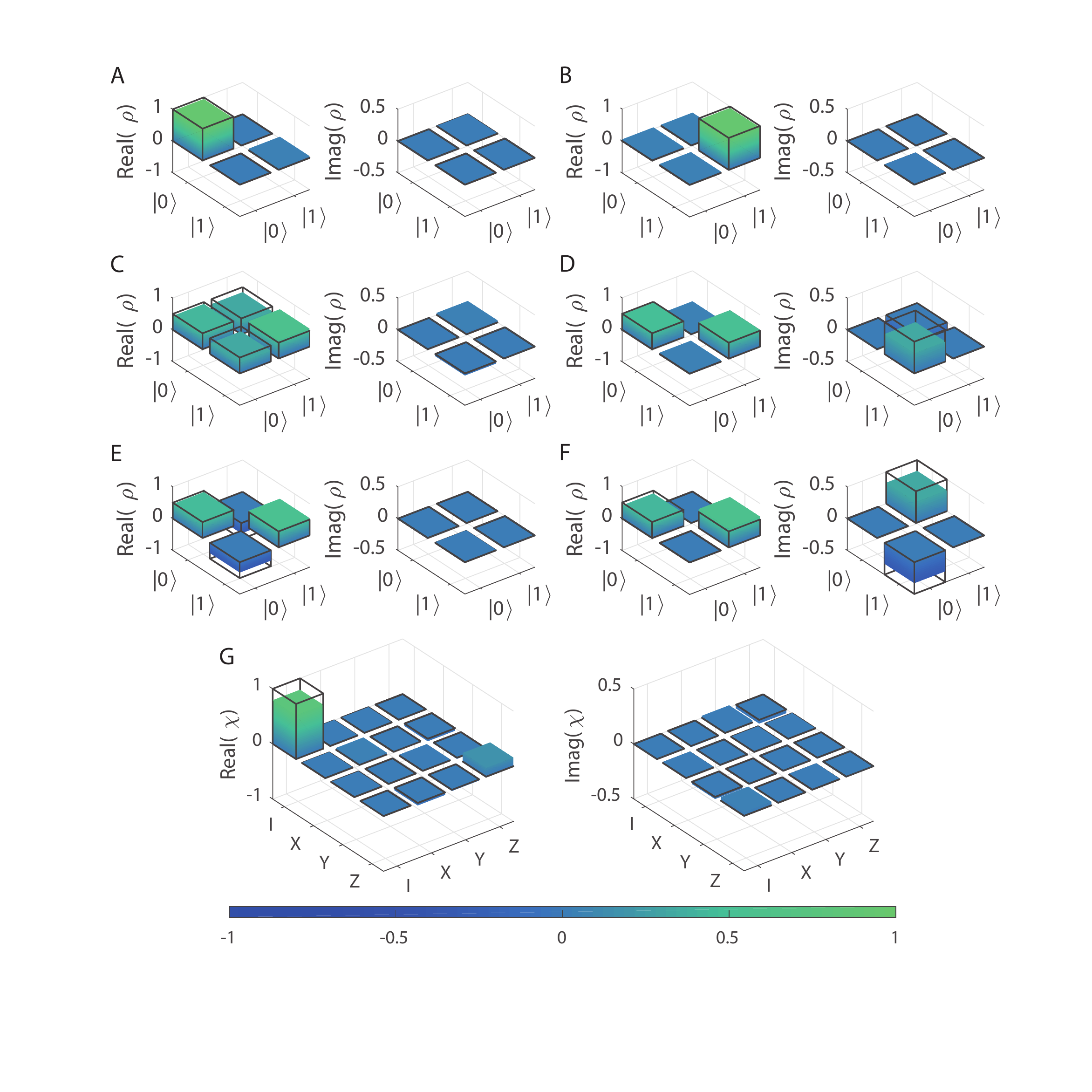}
	\caption{\textbf{Quantum state and process tomography of a successful reset.} {The free-evolution Hamiltonian is $I$, and the deterministic interaction is $U = (X\otimes Z+iY\otimes X)/\sqrt{2}$.} \textbf{A-F.} The density matries of target qubit after resetting with initialization to each axis of the Bloch sphere: Alphabetically, $|0\rangle$, $|1\rangle$,  $\ket{+}$, $\ket{i}$, $\ket{-}$ and $\ket{-i}$. The corresponding state fidelities are 0.946, 0.951, 0.840, 0.815, 0.823, and 0.829, respectively. \textbf{G.} The $\chi_i$ matrix of the resetting process determined from \textbf{A-D}. The process fidelity is 0.792$\pm$0.011. Solid lines correspond to ideal density matrices $\rho$ and the ideal $\chi_i$ matrix. The 95\% confidence intervals are estimated via bootstrapping (see Method for more details).}
	\label{density}
\end{figure*}

\begin{figure*}
	\centering
		\includegraphics[width=0.8\textwidth]{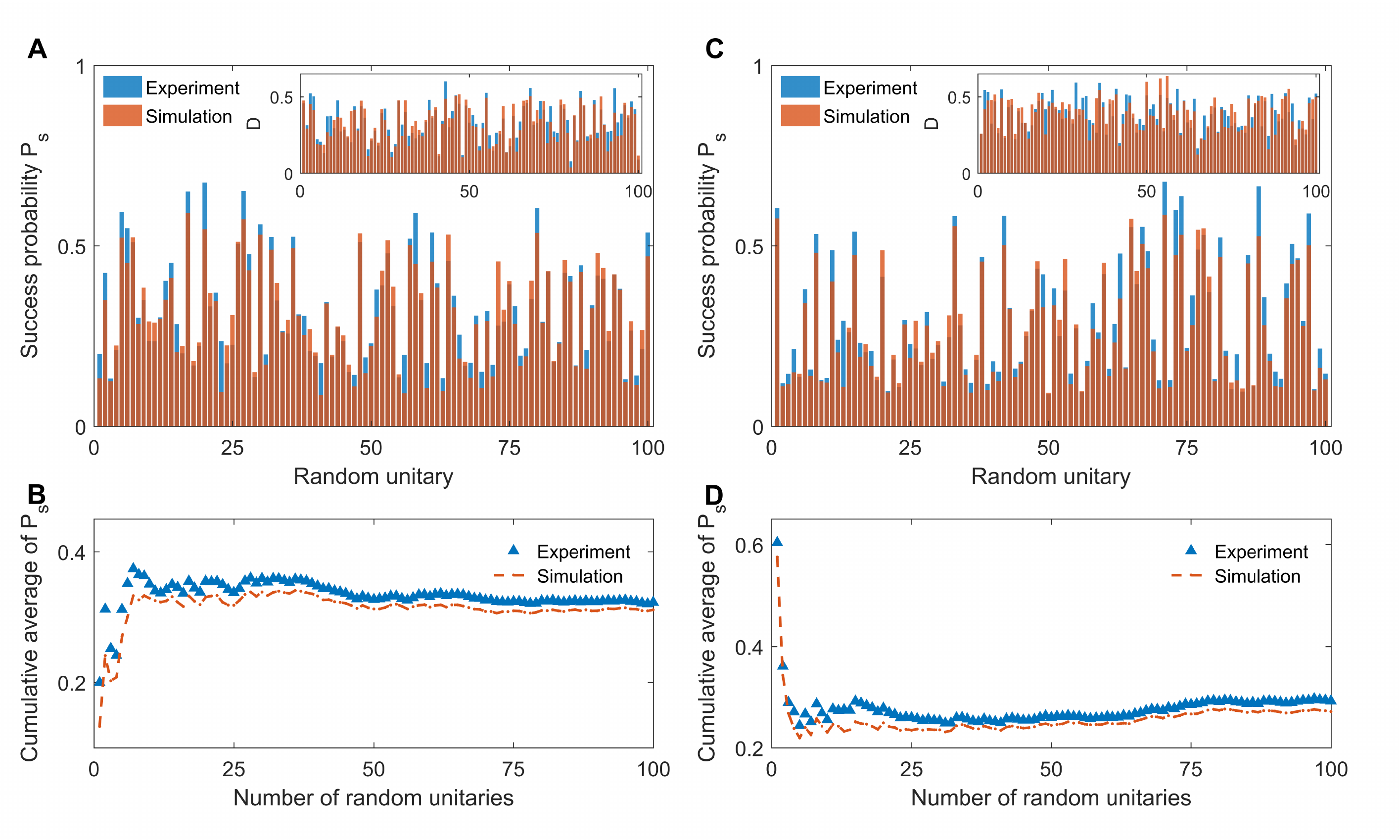}
	\caption{\textbf{Success probability of random unitaries for different initial states.} {The random unitaries are realized as $U_r$ shown in Fig. \ref{circuit}C, where 12 random single-qubit gates are applied before and after a $CZ$ gate.}  Experimental results are shown in blue; simulated values are in brown. In the simulation, decoherence effect is considered (see Method). As shown in case 3 of Table. \ref{case1}, the free-evolution Hamiltonian is $I$. In \textbf{A} and \textbf{B}, the initial state is \ket{-}, and in \textbf{C} and \textbf{D}, the initial state is \ket{1}. For these two different initial states, the random unitary sets are also different. \textbf{A} and \textbf{C.} Success probabilities for each of the 100 random unitaries; inset: the trace distances for each unitary. The experimental and simulated results are shown in blue and brown bars, respectively. \textbf{B} and \textbf{D.} Cumulative average of success probabilities. The experimental and simulated results are shown in blue triangles and brown dashed line, respectively.}
	\label{random}
\end{figure*}

\section{INTRODUCTION}
The tug of war between the natural but unknown evolution of a quantum system and control mechanisms to correct error introduced by such evolution is one of the most important technical challenges in the implementation of reliable quantum computers. Current protocols for removing unwanted evolution from a quantum system can be broadly classified into a few families. The widely used {refocusing} techniques such as spin echo~\cite{PhysRev.80.580}, dynamical decoupling~\cite{PhysRevLett.82.2417,zanardi,DUAN1999139} use fast control pulses to average out the effect of the unwanted evolution, effectively ``freezing'' the target system in time. The {universal refocusing} technique~\cite{1602.07963} is similar to refocusing, but has less assumptions on the target-control interaction Hamiltonian. A more recent technique~\cite{Murao2019} constructs a universal quantum circuit to probabilistically implement the exact inverse evolution of some quantum gate when only the dimension of the target and the number of permitted uses of the gate are known.

These techniques, however, rely on very different assumptions. None of the (universal) refocusing techniques are exact, and all of them require some assumptions about the dynamics. Spin echo and dynamical decoupling require the interaction between the target and the control not to be too strong. While universal refocusing works with stronger interactions, the trade-off between error $\epsilon$ and the number of ``controlled interactions'' $n$ scales as $n=O(\log^2{(1/\epsilon)})$. The protocol in~\cite{Murao2019} is exact, with a realistic probability of success which increases exponentially with the number of uses allowed for the gate to be inverted. But it does not directly remove an evolution, since the inverted gate must be subsequently applied to the target.

A recent protocol, {quantum resetting}~\cite{PhysRevX.8.031008}, combines some of the advantages of the existing protocols above:
\begin{enumerate}
\item Both the target and the interactions are {uncontrolled}. For the protocol to work, no information about the free evolution of the target or the interaction between the target and probes is needed, as long as these Hamiltonians are time-independent.
\item The target is {reset} to a state in the past when the protocol is successful, without the need of doing the exact inverse of its free evolution. 
\item It is {probabilistic} but {exact}, with realistic probabilities of success. For a qubit target and qubit probes, the probability of success is about 20\%.
\item Even if the protocol fails, it is possible to ``undo'' the failure by using the output of the failed protocol to increase its probability of success.
\end{enumerate}

The mathematical underpinning of the quantum resetting protocol is very different from the existing protocols above. Quantum resetting relies on {central matrix polynomials}~\cite{formanek_book}, which are polynomials such that when evaluated on arbitrary $n\times n$ matrices, the result is always a scalar matrix (i.e. proportional to the identity). In a quantum resetting protocol, the target system to be reset is allowed to evolve freely for time $t$ according to some unknown time-independent Hamiltonian $H_0$. After the free evolution, the target system interacts with a probe for time $t'$ governed by another unknown time-independent Hamiltonian $H_I$. When the free evolution/probe interaction cycle is completed for the last probe, the probes are measured. If the measurement is successful, then the effect of the free evolution/probe interaction is canceled, the target state has been reset to the past, before the first free evolution, otherwise the protocol fails. For the $\mathcal{W}_4$ protocol, the target and the probes are all qubits. The probes are prepared as 2 singlet states $\ket{\Psi^-}=\frac{1}{\sqrt{2}}(\ket{01}-\ket{10})$, and the measurement is a projection into the subspace $\mathcal{S}$ spanned by $\{\ket{0000},\;\ket{1111},\;\frac{1}{\sqrt{2}}(\ket{0011}+\ket{1100}),\;\frac{1}{2}(\ket{1000}+\ket{0100}+\ket{0010}+\ket{0001}),\;\frac{1}{2}(\ket{0111}+\ket{1011}+\ket{1101}+\ket{1110}),\;\frac{1}{2}(\ket{1010}+\ket{0101}+\ket{1001}+\ket{0110})\}$.

In our experiment, we tested the simplest non-trivial quantum resetting protocol: a 2D target interacting with four 2D probes, {known as the $\mathcal{W}_4$ protocol}~\cite{PhysRevX.8.031008}. 
After being set to an initial state, the target qubit interacts with each of the four probes, which form two pairs of entangled states. Then, measurement of the probes affects the target qubit, sending it back to its initial state with a given probability.


\section{RESULTS}
\textbf{Experimental implementation of the resetting protocol}\\
The general gate sequence of the quantum circuit used to implement the $\mathcal{W}_4$ protocol is pictured in Fig.~\ref{circuit}A. We divide the circuit into four parts: state preparation, free evolution, interaction and {tomographic readout}. Before the circuit begins, all qubits are {initialized} in the state $\ket{0}$. During state preparation, the gate $G_1$ is applied to the target qubit to bring it to $\ket{\psi(0)}$, and each pair of neighboring probes is set to the singlet state $|\Psi^-\rangle = \frac{\ket{01}-\ket{10}}{\sqrt{2}}$. After state preparation, we apply the gate $R$, which simulates the free evolution with Hamiltonian $\mathcal{H}_0=\sum\limits_{j=x,y,z}h_j\sigma_j$, where $\{\sigma_j\}$ are the Pauli operators and $h_j$ is the coupling strength on the $j$-axis. After the first $R$ gate is applied to the target, the interaction process started. The first probe interacts with the target via a bipartite unitary operator $U$, which varies according to the experimental case. This process of free evolution followed by target-probe interaction is repeated three more times on the target and different probes. We note that in this experiment, as there is no direct coupling between the next-nearest-neighboring qubits, the interaction operation between them is realized via the qubit between them. Take $Q_1$ and $Q_3$ as an example. To apply a bipartite unitary operator $U$ between them, we first apply $U$ between $Q_2$ and $Q_3$, and then apply a SWAP gate between $Q_2$ and $Q_1$. Both $U$ and SWAP gates are based on controlled-phase (CZ) gate. The SWAP gate is realized as the combination of single-qubit gates and three CZ gates. More details can be found in Method. Once the interaction process is complete, a five-qubit state tomography is performed to obtain the final state with density matrix $\rho_f$. A successful reset has occurred in the portion of the state that overlaps with the success subspace $\mathcal{S}$ given above.

Projecting the probe subspace onto this success subspace post-selects for a successful reset. The trace overlap between the measured state and the projected state is defined as the success probability, 
$P_s = \trace (\rho_f \rho_{ps}),$
where $\rho_{ps}$ is the density matrix of the projected state.
The reset state of the target qubit can be extracted from the projected state by tracing out the probes\cite{Nielsen2010},
$\rho = \trace_{probes}(\rho_{ps})$, 
where $\rho$ is the density matrix of the reset state.
Once the reset state has been obtained, we also evaluate the quality of the reset state. 
We use trace distance to identify the distinguishability between the reset state and the initial state of the target qubit, i.e., $D=\frac{1}{2}\trace{|\rho-\rho_0|}$, where $|X|=\sqrt{X^\dag X}$ and $\rho_0=\ketbra{\psi(0)}$.
Note that for the deterministic cases of our experiment, success subspace is reduced to the space spanned by the first three vectors in $\mathcal{S}$.

This protocol lies on the vanguard of what is currently experimentally feasible. Even for protocols with five qubits, correctly performing a quantum resetting protocol requires extremely high quality single- and double-qubit gates to model all possible interactions and free evolutions that make up the protocol. Quantum processors with superconducting qubits, which have undergone great progress over recent years, have reached a level of technical achievement that makes it possible to implement of the long sequences of arbitrary operations in a multi-qubit system~\cite{Gong2018,Roushan2017a,Chou2018,Hu2018,Wu2018,Xu2018,Kurpiers2018,Rosenberg2017,Kandala2017,Song2017,Versluis2017,Ofek2016,Corcoles2015,Kelly2015,Barends2014,Reed2012b,Yan2019,Arute2020,Lacroix2020}. In this experiment, we successfully implemented quantum circuits with up to totally 119 single-qubit gates and 12 entangling gates. The detail of the 47-layer circuits is shown in Method.


{To verify the $\mathcal{W}_4$ protocol,} we performed different variations of the {resetting} experiments, which we divide into three cases (Table~\ref{case1}). Case 1 {(Fig. \ref{proof})} tested interactions with a theoretical success probability of $1$, {i.e. deterministic interactions,} varying the initial target states and free evolutions. Case 2 {(Fig. \ref{density})} fully characterized the resetting process with six orthogonal initial states for the target. The final case {(Fig. \ref{random})} tested the success probability for random interactions, and compared them with the numerical predictions. {To characterize the success probability and the target state fidelity, we use tomographic readout in all cases to obtain the 5-qubit density matrix, and then project the state in the success subspace.}

\bigskip
\textbf{Test of the resetting protocol with deterministic interactions}\\
First we tested the protocol that the $U$ will deterministically reset target qubit in theory. To simulate effects from the physical evolution of a qubit, we varied the rotation around different axes, and measured $P_s$ and trace distances. As shown in Table~\ref{case1}, case 1a and 1c simulated the free evolution Hamiltonian $\mathcal{H}_0=h_z\sigma_z$ by applying a rotation gate $R_z(\varphi)$, corresponding to a rotation around $z$-axis through an angle $\varphi$. In case 1b the free evolution Hamiltonian was $\mathcal{H}_0=h_x\sigma_x$, rotating instead around the $x$-axis. Meanwhile, to observe the effect of free evolution sensitively, the target is initialized to a state orthogonal to the rotation axis of the free evolution, which in case 1a and 1c is $|-\rangle$, and in case 1b is $|1\rangle$. Furthermore, to test the protocol not only on pure states, but also on mixed states, case 1c allows the qubit to decohere for 1 $\mu$s in state preparation. The deterministic unitary depends on the form of free evolution Hamiltonian, thus was changed accordingly. Fig.~\ref{proof} illustrates the results of all three subcases for the rotation angle $\varphi=3\pi/8$, in which all fidelities of the reset states can be seen jumping above those without resetting. More results {verifying} the protocol for other rotation angles are listed in SM.

For case 1, we experimentally proved that the protocol can successfully reset the target with high fidelity using theoretically predicted deterministic unitaries. The success probabilities are not as high as theoretical prediction: for case 1a, we obtained $P_s = 0.544$. The results for case 1b and 1c are similar (see Supplementary Note 1 and Supplementary Table 1). We attribute this difference to the fidelity of the measured {5-qubit} state, which was found to be $0.399$ {in comparison with an ideal state}. We find the short dephasing time could be the main reason which limited the fidelity. In our simulation, if considering the decoherence, the 5-qubit state fidelity will drop to a similar value ~0.386 (See the Method). In context of these long quantum circuits, {trace distances of reset target, as shown in Fig. \ref{proof}D,H,L, }in the range of $0.21-0.28$ really stand out. 


\bigskip
\textbf{Characterization of the protocol with quantum process tomography}\\
Once we confirmed that the protocol can reset the target qubit, we decided to characterize the resetting process more closely. Setting $R = I$ and $U = (X\otimes Z+iY\otimes X)/\sqrt{2}$ (shown in Fig.~\ref{circuit}B), we initialized the target qubit to the six axes of the Bloch sphere and performed quantum process tomography. The density matrices $\rho$ of the reset target obtained with state tomography has significant variations in fidelity depending on the initialization (Figure\ref{density}A-F). The state fidelity is $F=|\trace(\sqrt{\sqrt{\rho_f} \rho_{ideal} \sqrt{\rho_f}})|^2$. The states $|0\rangle$ and $|1\rangle$ are not sensitive to dephasing, and have higher reset fidelities -- close to $0.95$. But the four other initializations are located on the equator of the Bloch sphere, so they are sensitive to dephasing and accordingly, have lower fidelites, ranging from $0.81 - 0.84$. These different initialization are important because they can be used to fully characterize the resetting process. By combining final states $\ket{0}$, $\ket{1}$, $\ket{+}$ and $\ket{i}$, we can obtain the $\chi$ matrix (Fig. \ref{density}G) with quantum process tomography \cite{Nielsen2010}. A completely positive and trace-preserving (CPTP) projection \cite{Knee2018} is used to guarantee a physical estimate of the $\chi$ matrix. We define the process fidelity as the trace overlap between the ideal process $\chi_i$, which only contains the identity operation $I$, and the measured $\chi$, as $\mathcal{F}_{\chi} = Tr(\chi_i\chi)$, {and} is determined to be above 0.79. The comparison between the reset fidelities of phase-sensitive and phase-insensitive initial states shows the important role dephasing plays in our experiment, leaving room for further improvement.

\bigskip
\textbf{Implementation of the resetting protocol with random interactions}\\
The most remarkable advantage of this resetting protocol is that the interaction need not to be controlled or known. {Our interpretation of the word `known' simply means that} the interaction dynamics can not be adjusted according to the free evolution of the target system. To simulate these sorts of situations, we investigate the effects of random target-probe interactions. Specifically, we tested the success probability of random unitaries ($U_r$), generated by rotating the target qubit and its interaction probe before and after a $CZ$ gate (Fig. \ref{circuit}C). Each random rotation is implemented as a sequence of $R_z(\alpha_i)$, $R_y(\beta_i)$, and $R_z(\gamma_i)$ gates, with angles $\alpha_i$, $\beta_i$, and $\gamma_i$ all chosen randomly. As shown in Table~\ref{case1} case 3, we set $R=I$ and $\ket{\psi(0)}=\ket{1}$ and \ket{-}, respectively, and tested two sets of 100 different random unitaries. {To compare the experimental and theoretical results, we numerically simulated the circuit with decoherence and calculated the corresponding success probability and state fidelity.} {Experimental} success probabilities for the random unitaries are in good agreement with numerical simulation for both initial states (Fig.~\ref{random}). When the results for the random unitaries are combined, the cumulative average of success probability converges towards $0.323$ and $0.292$ for these two initial states, which is close to the simulated value of $0.312$ and $0.271$, respectively. The difference of experimental cumulative average comes from limited number of random unitaries. In our simulation, when the number of random unitaries increases to 5000, the cumulative average of success probability converges toward $0.290$ for both initial states. The experimental average trace distances are $0.334$ and $0.391$, respectively. We attribute the poor trace distances mainly to decoherence, especially dephasing. In our simulation, after increasing the dephasing time to be the same as the relaxation time, the average trace distances can be decreased from $0.322$ and $0.390$, to $0.155$ and $0.205$, respectively. {Similar to} our other results, we expect these blemishes to become less pronounced as the quality of quantum processors is improved.


\section{DISCUSSION}
We have successfully {verified} the quantum resetting protocol for known and unknown interactions. Even when the interactions are not known, we still have an average success probability of about $0.3$. This probability can be significantly improved by an `undoing failure' protocol presented by Navascu\'es\cite{PhysRevX.8.031008}. Upon failure, it is possible to send more probes to interact with the target, and measure the new probes for another chance of a successful reset. {Although practical difficulties in implementing additional layers of circuit to correct failed resets may outweigh the potential benefits, since the ``undoing failure'' protocol may not increase the fidelity of the reset state.}

Another result of Navascu\'es \cite{PhysRevX.8.031008}, is that the resetting protocol can reset a target system of any dimension. In a photonic system, our colleagues have demonstrated that a qubit can be reset to its past entangled state\cite{Li2019}, and also here we showed that a mixed state can be reset, giving the experimental {verification} that the protocol can work on a target qubit which is a part of higher-dimensional systems. Given the speed of progress with superconducting processors, it is expected that the realization of resetting higher-dimensional systems is achievable in the near term, opening the door for applications in quantum memory\cite{terhal2015quantum}. \blue{It is also notable that even though the free evolution/target-probe interactions must be unitary, it is possible to model an open system by purifying the target and the dynamics. Thus, it is possible that quantum resetting with a higher-dimensional target can be used to model the open dynamics of its lower-dimensional subspace, making it an interesting alternative to quantum error correction~\cite{campbell2017roads}. Even though quantum resetting protocols exist for targets of any dimension, currently only the ones with qubit and qutrit targets can be written down explicitly. Another challenge to implement $d$-dimensional quantum resetting protocols is the depth of the circuit, with the number of probes scaling as $O(d^3)$, thus requiring extremely good quantum gates and long coherence time. However, unlike quantum error correction, quantum resetting protocols do not require a large number of qubits. To explore this direction, theoretical tools for easily finding deterministic interactions in higher dimensions are urgently needed.} We expect that theoretical and experimental development of this protocol will have great potential to advance many areas of quantum information processing.

\bibliographystyle{naturemag}

\begin{thebibliography}{10}
	\expandafter\ifx\csname url\endcsname\relax
	\def\url#1{\texttt{#1}}\fi
	\expandafter\ifx\csname bibnamefont\endcsname\relax
	\def\bibnamefont#1{#1}\fi
	\expandafter\ifx\csname urlprefix\endcsname\relax\def\urlprefix{URL }\fi
	\providecommand{\bibinfo}[2]{#2}
	\providecommand{\eprint}[2][]{\url{#2}}
	
	\bibitem{PhysRev.80.580}
	\bibinfo{author}{Hahn, E.~L.}
	\newblock \bibinfo{title}{{Spin Echoes}}.
	\newblock \textit{\bibinfo{journal}{Phys. Rev.}} \textbf{\bibinfo{volume}{80}},
	\bibinfo{pages}{580--594} (\bibinfo{year}{1950}).
	
	\bibitem{PhysRevLett.82.2417}
	\bibinfo{author}{Viola, L.}, \bibinfo{author}{Knill, E.} \&
	\bibinfo{author}{Lloyd, S.}
	\newblock \bibinfo{title}{{Dynamical Decoupling of Open Quantum Systems}}.
	\newblock \textit{\bibinfo{journal}{Phys. Rev. Lett.}}
	\textbf{\bibinfo{volume}{82}}, \bibinfo{pages}{2417--2421}
	(\bibinfo{year}{1999}).
	
	\bibitem{zanardi}
	\bibinfo{author}{Zanardi, P.}
	\newblock \bibinfo{title}{{Symmetrizing evolutions}}.
	\newblock \textit{\bibinfo{journal}{Phys. Lett. A}}
	\textbf{\bibinfo{volume}{258}}, \bibinfo{pages}{77 -- 82}
	(\bibinfo{year}{1999}).
	
	\bibitem{DUAN1999139}
	\bibinfo{author}{Duan, L.-M.} \& \bibinfo{author}{Guo, G.-C.}
	\newblock \bibinfo{title}{Suppressing environmental noise in quantum
		computation through pulse control}.
	\newblock \textit{\bibinfo{journal}{Phys. Lett. A}}
	\textbf{\bibinfo{volume}{261}}, \bibinfo{pages}{139 -- 144}
	(\bibinfo{year}{1999}).
	
	\bibitem{1602.07963}
	\bibinfo{author}{Sardharwalla, I. S.~B.}, \bibinfo{author}{Cubitt, T.~S.},
	\bibinfo{author}{Harrow, A.~W.} \& \bibinfo{author}{Linden, N.}
	\newblock \bibinfo{title}{{Universal Refocusing of Systematic Quantum Noise}}.
	Preprint at \url{https://arxiv.org/abs/1602.07963} (\bibinfo{year}{2016}).
	
	\bibitem{Murao2019}
	\bibinfo{author}{Quintino, M.~T.}, \bibinfo{author}{Dong, Q.},
	\bibinfo{author}{Shimbo, A.}, \bibinfo{author}{Soeda, A.} \&
	\bibinfo{author}{Murao, M.}
	\newblock \bibinfo{title}{Reversing unknown quantum transformations: Universal
		quantum circuit for inverting general unitary operations}.
	\newblock \textit{\bibinfo{journal}{Phys. Rev. Lett.}}
	\textbf{\bibinfo{volume}{123}}, \bibinfo{pages}{210502}
	(\bibinfo{year}{2019}).
	
	\bibitem{PhysRevX.8.031008}
	\bibinfo{author}{Navascu\'es, M.}
	\newblock \bibinfo{title}{{Resetting Uncontrolled Quantum Systems}}.
	\newblock \textit{\bibinfo{journal}{Phys. Rev. X}} \textbf{\bibinfo{volume}{8}},
	\bibinfo{pages}{031008} (\bibinfo{year}{2018}).
	
	\bibitem{formanek_book}
	\bibinfo{author}{Formanek, E.}
	\newblock \textit{\bibinfo{title}{The Polynomial Identities and Invariants of
			$n\times n$ Matrices}}.
	\newblock No.~\bibinfo{number}{78} in \bibinfo{series}{Regional Conference
		Series in Mathematics} (\bibinfo{publisher}{American Mathematical Society},
	\bibinfo{year}{1991}).
	
	\bibitem{Nielsen2010}
	\bibinfo{author}{Nielsen, M.~A.} \& \bibinfo{author}{Chuang, I.~L.}
	\newblock \textit{\bibinfo{title}{{Quantum computation and quantum information}}}
	(\bibinfo{publisher}{Cambridge University Press},
	\bibinfo{address}{Cambridge}, \bibinfo{year}{2010}).
	
	\bibitem{Gong2018}
	\bibinfo{author}{Gong, M.}, \bibnamefont{et~al.}
	\newblock \bibinfo{title}{{Genuine 12-Qubit Entanglement on a Superconducting
			Quantum Processor}}.
	\newblock \textit{\bibinfo{journal}{Phys. Rev. Lett.}}
	\textbf{\bibinfo{volume}{122}}, \bibinfo{pages}{110501}
	(\bibinfo{year}{2019}).
	
	\bibitem{Roushan2017a}
	\bibinfo{author}{Roushan, P.}, \bibnamefont{et~al.}
	\newblock \bibinfo{title}{{Spectroscopic signatures of localization with
			interacting photons in superconducting qubits}}.
	\newblock \textit{\bibinfo{journal}{Science}} \textbf{\bibinfo{volume}{358}},
	\bibinfo{pages}{1175--1179} (\bibinfo{year}{2017}).
	
	\bibitem{Chou2018}
	\bibinfo{author}{Chou, K.~S.}, \bibnamefont{et~al.}
	\newblock \bibinfo{title}{{Deterministic teleportation of a quantum gate
			between two logical qubits}}.
	\newblock \textit{\bibinfo{journal}{Nat.}} \textbf{\bibinfo{volume}{561}},
	\bibinfo{pages}{368--373} (\bibinfo{year}{2018}).
	
	\bibitem{Hu2018}
	\bibinfo{author}{Hu, L.}, \bibnamefont{et~al.}
	\newblock \bibinfo{title}{{Demonstration of quantum error correction and
			universal gate set on a binomial bosonic logical qubit}}. 
	\newblock \textit{\bibinfo{journal}{Nat. Phys.}}
	\textbf{\bibinfo{volume}{15}}, \bibinfo{pages}{503-508} 
	(\bibinfo{year}{2018}).
	
	\bibitem{Wu2018}
	\bibinfo{author}{Wu, Y.}, \bibnamefont{et~al.}
	\newblock \bibinfo{title}{{An efficient and compact switch for quantum
			circuits}}.
	\newblock \textit{\bibinfo{journal}{npj Quantum Inf.}}
	\textbf{\bibinfo{volume}{4}}, \bibinfo{pages}{50} (\bibinfo{year}{2018}).
	
	\bibitem{Xu2018}
	\bibinfo{author}{Xu, K.}, \bibnamefont{et~al.}
	\newblock \bibinfo{title}{{Emulating Many-Body Localization with a
			Superconducting Quantum Processor}}.
	\newblock \textit{\bibinfo{journal}{Phys. Rev. Lett.}}
	\textbf{\bibinfo{volume}{120}}, \bibinfo{pages}{050507}
	(\bibinfo{year}{2018}).
	
	\bibitem{Kurpiers2018}
	\bibinfo{author}{Kurpiers, P.}, \bibnamefont{et~al.}
	\newblock \bibinfo{title}{{Deterministic quantum state transfer and remote
			entanglement using microwave photons}}.
	\newblock \textit{\bibinfo{journal}{Nat.}} \textbf{\bibinfo{volume}{558}},
	\bibinfo{pages}{264--267} (\bibinfo{year}{2018}).
	
	\bibitem{Rosenberg2017}
	\bibinfo{author}{Rosenberg, D.}, \bibnamefont{et~al.}
	\newblock \bibinfo{title}{{3D integrated superconducting qubits}}.
	\newblock \textit{\bibinfo{journal}{npj Quantum Inf.}}
	\textbf{\bibinfo{volume}{3}}, \bibinfo{pages}{42} (\bibinfo{year}{2017}).
	
	\bibitem{Kandala2017}
	\bibinfo{author}{Kandala, A.}, \bibnamefont{et~al.}
	\newblock \bibinfo{title}{{Hardware-efficient variational quantum eigensolver
			for small molecules and quantum magnets}}.
	\newblock \textit{\bibinfo{journal}{Nat.}} \textbf{\bibinfo{volume}{549}},
	\bibinfo{pages}{242--246} (\bibinfo{year}{2017}).
	
	\bibitem{Song2017}
	\bibinfo{author}{Song, C.}, \bibnamefont{et~al.}
	\newblock \bibinfo{title}{{10-Qubit Entanglement and Parallel Logic Operations
			with a Superconducting Circuit}}.
	\newblock \textit{\bibinfo{journal}{Phys. Rev. Lett.}}
	\textbf{\bibinfo{volume}{119}}, \bibinfo{pages}{180511}
	(\bibinfo{year}{2017}).
	
	\bibitem{Versluis2017}
	\bibinfo{author}{Versluis, R.}, \bibnamefont{et~al.}
	\newblock \bibinfo{title}{{Scalable Quantum Circuit and Control for a
			Superconducting Surface Code}}.
	\newblock \textit{\bibinfo{journal}{Phys. Rev. Appl.}}
	\textbf{\bibinfo{volume}{8}} (\bibinfo{year}{2017}).
	
	\bibitem{Ofek2016}
	\bibinfo{author}{Ofek, N.}, \bibnamefont{et~al.}
	\newblock \bibinfo{title}{{Extending the lifetime of a quantum bit with error
			correction in superconducting circuits}}.
	\newblock \textit{\bibinfo{journal}{Nat.}} \textbf{\bibinfo{volume}{536}},
	\bibinfo{pages}{441--445} (\bibinfo{year}{2016}).
	
	\bibitem{Corcoles2015}
	\bibinfo{author}{C{\'{o}}rcoles, A.~D.}, \bibnamefont{et~al.}
	\newblock \bibinfo{title}{{Demonstration of a quantum error detection code
			using a square lattice of four superconducting qubits}}.
	\newblock \textit{\bibinfo{journal}{Nat. Commun.}}
	\textbf{\bibinfo{volume}{6}} (\bibinfo{year}{2015}).
	
	\bibitem{Kelly2015}
	\bibinfo{author}{Kelly, J.}, \bibnamefont{et~al.}
	\newblock \bibinfo{title}{{State preservation by repetitive error detection in
			a superconducting quantum circuit}}.
	\newblock \textit{\bibinfo{journal}{Nat.}} \textbf{\bibinfo{volume}{519}},
	\bibinfo{pages}{66--69} (\bibinfo{year}{2015}).
	
	\bibitem{Barends2014}
	\bibinfo{author}{Barends, R.}, \bibnamefont{et~al.}
	\newblock \bibinfo{title}{{Superconducting quantum circuits at the surface code
			threshold for fault tolerance}}.
	\newblock \textit{\bibinfo{journal}{Nat.}} \textbf{\bibinfo{volume}{508}},
	\bibinfo{pages}{500--503} (\bibinfo{year}{2014}).
	
	\bibitem{Reed2012b}
	\bibinfo{author}{Reed, M.~D.}, \bibnamefont{et~al.}
	\newblock \bibinfo{title}{{Realization of three-qubit quantum error correction
			with superconducting circuits}}.
	\newblock \textit{\bibinfo{journal}{Nat.}} \textbf{\bibinfo{volume}{482}},
	\bibinfo{pages}{382--385} (\bibinfo{year}{2012}).
	
	\bibitem{Yan2019}
	\bibinfo{author}{Yan, Z.}, \bibnamefont{et~al.}
	\newblock \bibinfo{title}{{Strongly correlated quantum walks with a 12-qubit
			superconducting processor}}.
	\newblock \textit{\bibinfo{journal}{Science}} \textbf{\bibinfo{volume}{364}},
	\bibinfo{pages}{753-756} (\bibinfo{year}{2019}).
	
	\bibitem{Arute2020}
	\bibinfo{author}{Arute, F.}, \bibnamefont{et~al.}
	\newblock \bibinfo{title}{{Quantum Approximate Optimization of Non-Planar Graph
			Problems on a Planar Superconducting Processor}}.
	Preprint at \url{http://arxiv.org/abs/2004.04197} (\bibinfo{year}{2020}).
	
	\bibitem{Lacroix2020}
	\bibinfo{author}{Lacroix, N.}, \bibnamefont{et~al.}
	\newblock \bibinfo{title}{{Improving the Performance of Deep Quantum
			Optimization Algorithms with Continuous Gate Sets}}.
	Preprint at \url{http://arxiv.org/abs/2005.05275}  (\bibinfo{year}{2020}).
	
	\bibitem{Knee2018}
	\bibinfo{author}{Knee, G.~C.}, \bibinfo{author}{Bolduc, E.},
	\bibinfo{author}{Leach, J.} \& \bibinfo{author}{Gauger, E.~M.}
	\newblock \bibinfo{title}{{Quantum process tomography via completely positive
			and trace-preserving projection}}.
	\newblock \textit{\bibinfo{journal}{Phys. Rev. A}}
	\textbf{\bibinfo{volume}{98}} (\bibinfo{year}{2018}).
	
	\bibitem{Li2019}
	\bibinfo{author}{Li, Z.-D.}, \bibnamefont{et~al.}
	\newblock \bibinfo{title}{{Photonic realization of quantum resetting}}.
	\newblock \textit{\bibinfo{journal}{Optica}} \textbf{\bibinfo{volume}{7}},
	\bibinfo{pages}{766} (\bibinfo{year}{2020}).
	
	\bibitem{terhal2015quantum}
	\bibinfo{author}{Terhal, B.~M.}
	\newblock \bibinfo{title}{Quantum error correction for quantum memories}.
	\newblock \textit{\bibinfo{journal}{Rev. Mod. Phys.}}
	\textbf{\bibinfo{volume}{87}}, \bibinfo{pages}{307} (\bibinfo{year}{2015}).
	
	\bibitem{campbell2017roads}
	\bibinfo{author}{Campbell, E.~T.}, \bibinfo{author}{Terhal, B.~M.} \&
	\bibinfo{author}{Vuillot, C.}
	\newblock \bibinfo{title}{Roads towards fault-tolerant universal quantum
		computation}.
	\newblock \textit{\bibinfo{journal}{Nat.}} \textbf{\bibinfo{volume}{549}},
	\bibinfo{pages}{172} (\bibinfo{year}{2017}).
	
	\bibitem{Barends2013}
	\bibinfo{author}{Barends, R.}, \bibnamefont{et~al.}
	\newblock \bibinfo{title}{{Coherent josephson qubit suitable for scalable
			quantum integrated circuits}}.
	\newblock \textit{\bibinfo{journal}{Phys. Rev. Lett.}}
	\textbf{\bibinfo{volume}{111}}, \bibinfo{pages}{1--6} (\bibinfo{year}{2013}).
	
	\bibitem{Koch2007}
	\bibinfo{author}{Koch, J.}, \bibnamefont{et~al.}
	\newblock \bibinfo{title}{{Charge-insensitive qubit design derived from the
			Cooper pair box}}.
	\newblock \textit{\bibinfo{journal}{Phys. Rev. A}}
	\textbf{\bibinfo{volume}{76}}, \bibinfo{pages}{1--19} (\bibinfo{year}{2007}).
	
	\bibitem{Gong2019}
	\bibinfo{author}{Gong, M.}, \bibnamefont{et~al.}
	\newblock \bibinfo{title}{{Experimental verification of five-qubit quantum
			error correction with superconducting qubits}}. Preprint at \url{https://arxiv.org/abs/1907.04507}
	(\bibinfo{year}{2019}).
	
	\bibitem{Steffen2006c}
	\bibinfo{author}{Steffen, M.}, \bibnamefont{et~al.}
	\newblock \bibinfo{title}{{Measurement of the entanglement of two
			superconducting qubits via state tomography}}.
	\newblock \textit{\bibinfo{journal}{Science}} \textbf{\bibinfo{volume}{313}},
	\bibinfo{pages}{1423--1425} (\bibinfo{year}{2006}).
	
	\bibitem{Zheng2017}
	\bibinfo{author}{Zheng, Y.}, \bibnamefont{et~al.}
	\newblock \bibinfo{title}{{Solving Systems of Linear Equations with a
			Superconducting Quantum Processor}}.
	\newblock \textit{\bibinfo{journal}{Phys. Rev. Lett.}}
	\textbf{\bibinfo{volume}{118}}, \bibinfo{pages}{210504}
	(\bibinfo{year}{2017}).
	
	\bibitem{Dicarlo2009}
	\bibinfo{author}{Dicarlo, L.}, \bibnamefont{et~al.}
	\newblock \bibinfo{title}{{Demonstration of two-qubit algorithms with a
			superconducting quantum processor}}.
	\newblock \textit{\bibinfo{journal}{Nat.}} \textbf{\bibinfo{volume}{460}},
	\bibinfo{pages}{240--244} (\bibinfo{year}{2009}).
	
	\bibitem{Martinis2014}
	\bibinfo{author}{Martinis, J.~M.} \& \bibinfo{author}{Geller, M.~R.}
	\newblock \bibinfo{title}{{Fast adiabatic qubit gates using only $\sigma_z$
			control}}.
	\newblock \textit{\bibinfo{journal}{Phys. Rev. A}}
	\textbf{\bibinfo{volume}{90}}, \bibinfo{pages}{022307}
	(\bibinfo{year}{2014}).
	
	\bibitem{Kelly2014}
	\bibinfo{author}{Kelly, J.}, \bibnamefont{et~al.}
	\newblock \bibinfo{title}{{Optimal quantum control using randomized
			benchmarking}}.
	\newblock \textit{\bibinfo{journal}{Phys. Rev. Lett.}}
	\textbf{\bibinfo{volume}{112}}, \bibinfo{pages}{240504}
	(\bibinfo{year}{2014}).
	
	\bibitem{Neill2018}
	\bibinfo{author}{Neill, C.}, \bibnamefont{et~al.}
	\newblock \bibinfo{title}{{A blueprint for demonstrating quantum supremacy with
			superconducting qubits}}.
	\newblock \textit{\bibinfo{journal}{Science}} \textbf{\bibinfo{volume}{360}},
	\bibinfo{pages}{195--199} (\bibinfo{year}{2018}).
	
	\bibitem{Johnson2011d}
	\bibinfo{author}{Johnson, B.~R.}
	\newblock \textit{\bibinfo{title}{{Controlling Photons in Superconducting Electrical
				Circuits}}}.
	\newblock ({\bibinfo{school}{Yale University}},  \bibinfo{year}{2011}).
	
	\bibitem{Gustavsson2012}
	\bibinfo{author}{Gustavsson, S.}, \bibnamefont{et~al.}
	\newblock \bibinfo{title}{{Dynamical decoupling and dephasing in interacting
			two-level systems}}.
	\newblock \textit{\bibinfo{journal}{Phys. Rev. Lett.}}
	\textbf{\bibinfo{volume}{109}}, \bibinfo{pages}{1--5} (\bibinfo{year}{2012}).
	
	\bibitem{Martinis2003}
	\bibinfo{author}{Martinis, J.~M.}, \bibinfo{author}{Nam, S.},
	\bibinfo{author}{Aumentado, J.}, \bibinfo{author}{Lang, K.~M.} \&
	\bibinfo{author}{Urbina, C.}
	\newblock \bibinfo{title}{{Decoherence of a superconducting qubit due to bias
			noise}}.
	\newblock \textit{\bibinfo{journal}{Phys. Rev. B}}
	\textbf{\bibinfo{volume}{67}}, \bibinfo{pages}{94510} (\bibinfo{year}{2003}).
	
	\bibitem{Vion2003}
	\bibinfo{author}{Vion, D.}, \bibnamefont{et~al.}
	\newblock \bibinfo{title}{{Rabi oscillations, Ramsey fringes and spin echoes in
			an electrical circuit}}.
	\newblock \textit{\bibinfo{journal}{Fortschritte der Phys.}}
	\textbf{\bibinfo{volume}{51}}, \bibinfo{pages}{462--468}
	(\bibinfo{year}{2003}).
	
	
\end{thebibliography}

\section{METHOD}

\textbf{Experimental wiring setup and device}

\begin{figure*}[!htbp]
	\centering
	\includegraphics[width=0.8\textwidth]{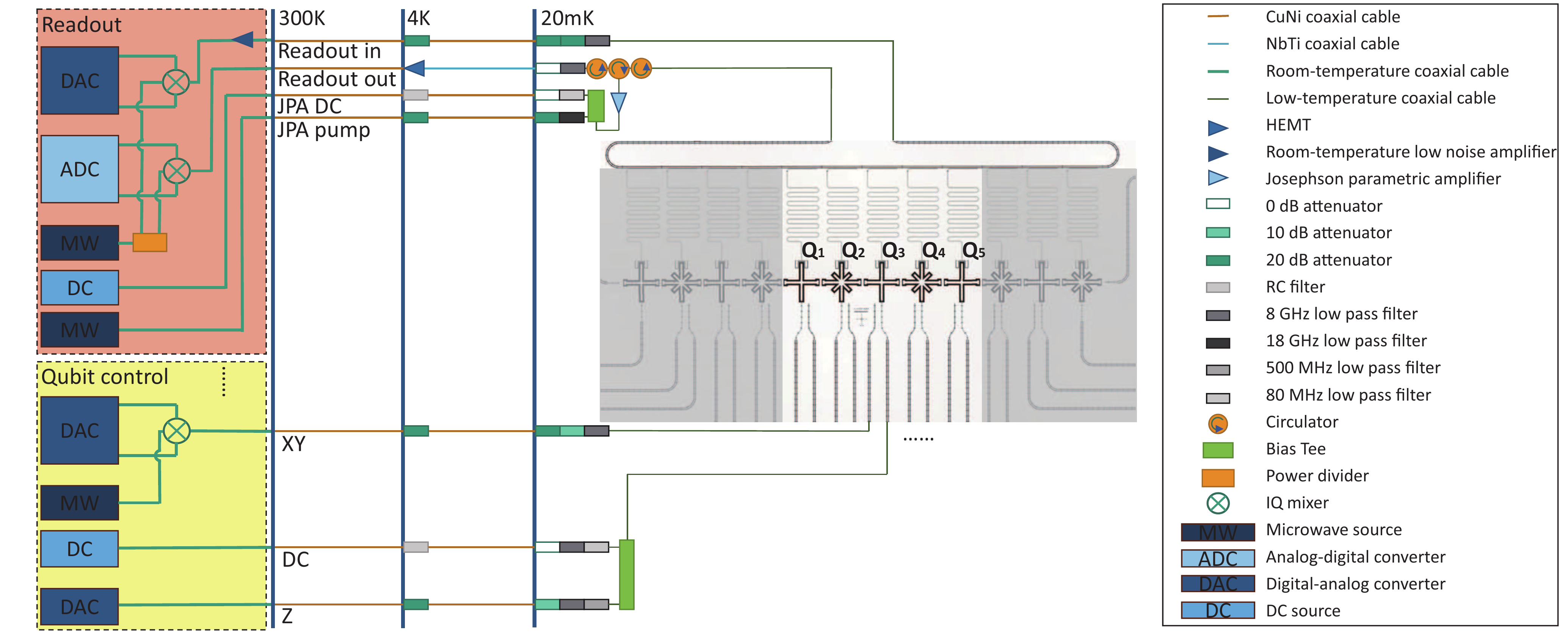}
	\caption{\textbf{Experimental wiring setup and false colored optical image of the qubit device.} There are five adjacent qubits used in this experiment. To perform the state manipulation and detection, five sets of qubit control and one set of readout units are used. For the control of each qubit, there are one microwave (XY) control and one flux bias. We use two digital-analog converter (DAC) channels and one microwave source to modulate the control signal via an IQ mixer. The XY signal is then wire-bonded to the XY control line, which is capacitively coupled to the qubit. The low speed and high speed bias (Z) signal is generated by a DC source and a DAC channel, and then combined together at the mixing chamber (MXC) plate via a bias-Tee. The Z signal is also wire-bonded to the Z control line, which is inductively coupled to the qubit. For state detection, the readout signal is modulated by two DAC channels and one microwave source via an IQ mixer. The signal is attenuated for 60 dB and then connected to the transmission line of the device. Then, the readout signal goes to the array of circulators. On the second circulator, the signal is amplified by the Josephson parametric amplifier (JPA), which is biased by the DC source and driven by the microwave source. The readout signal is then amplified by a HEMT at the 4K plate and a low-noise amplifier at room temperature, and finally demodulated by an IQ mixer and captured by the analog-digital converter (ADC). We use frequency multiplexing method to simultaneously readout the states of all qubits. All room temperature electronic instruments are controlled by a computer via 1Gb Ethernet.}
	\label{S2}
\end{figure*}

The superconducting processor used in this work is a 12-qubits processor\cite{Gong2018,Yan2019}, as shown in Fig. \ref{S2}. The twelve qubits are arranged in a 1-D lattice. Each qubit is an Xmon variant \cite{Barends2013} of transmon qubit \cite{Koch2007}. The qubits couple to their nearest-neighbor qubits via a fixed capacitor, inducing a constant coupling strength around ~11.5 MHz. For each qubit, there are one inductively coupled flux control line and one capacitively coupled microwave control line to enable the fully control of the quantum state. After the control operations, we simultaneously read out the state of all qubits via their dispersively coupled resonators. We chose five adjacent qubits \cite{Gong2019} to perform the present experiment, labeled as $Q_1$ to $Q_5$ in Fig. \ref{S2}. The device is mounted under the mixing chamber plate of a dilution refrigerator, for which the base temperature is below 12 mK. The experimental wiring setup from room temperature to low temperature is shown in Fig. \ref{S2}, too. Table. \ref{performance} shows the performance of the qubits in our experiment.

\begin{figure*}[!htbp]
	\centering
	\includegraphics[width=0.8\textwidth]{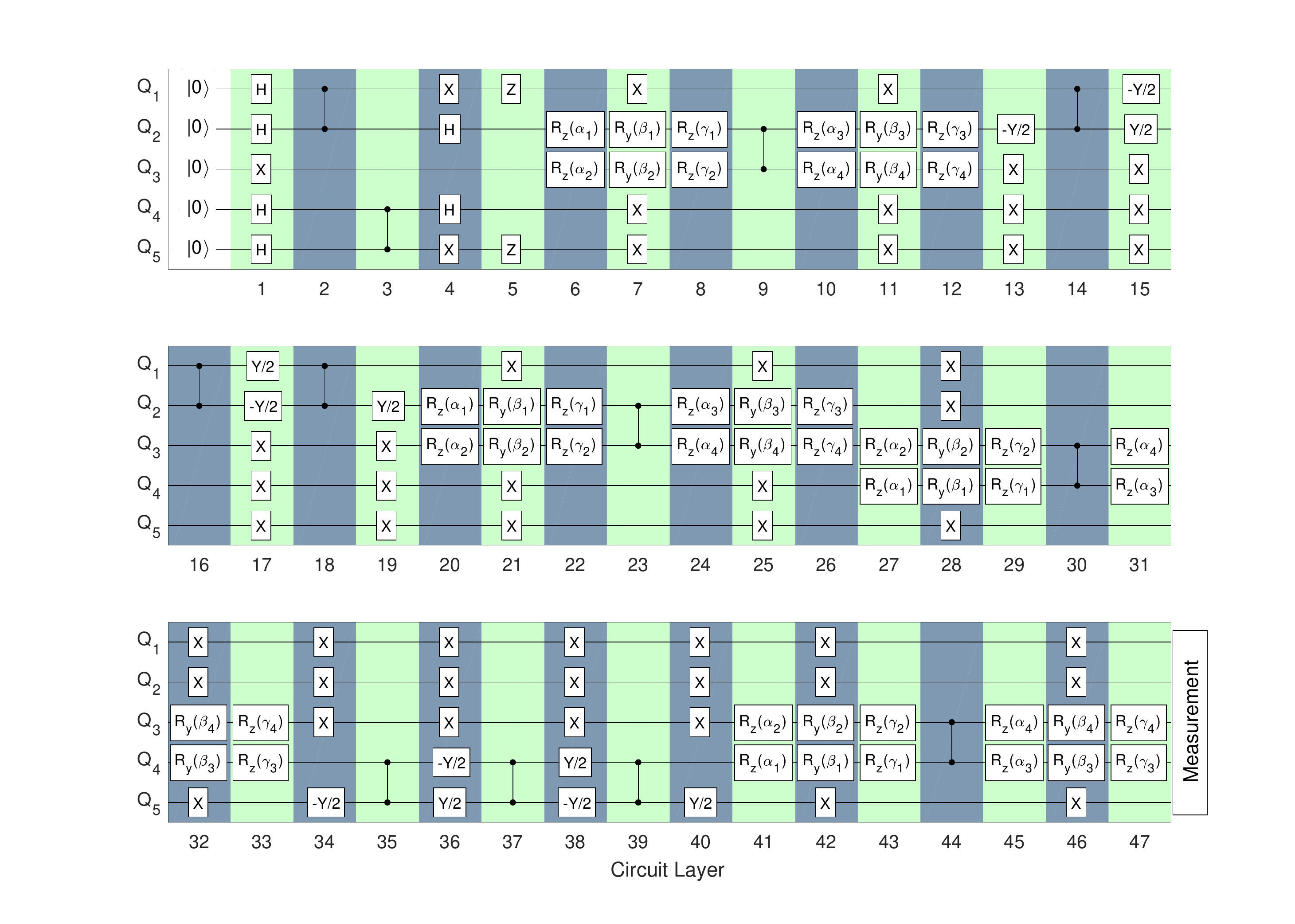}
	\caption{\textbf{Gate sequences for the test of random unitaries.} The initial state of the target qubit is \ket{1}. The total depth is 47, including 12 double-qubit gates and 119 single-qubit gates.}
	\label{S1}
\end{figure*}

\textbf{Gate implementation}

Only single-qubit gates and controlled-phase (CZ) gates are used in our experiment. Single-qubit gates are implemented as microwave pulses. We realize CZ gates by implementing DC wave sequences on two neighbouring qubits to tune the $|11\rangle$ state close to the avoided crossing generated by the states $|11\rangle$ and $|02\rangle$ following a "fast adiabatic" trajectory \cite{Dicarlo2009, Martinis2014, Barends2014, Kelly2014}.

CZ gates can only be implemented on neighbouring qubits, so to generate interactions between distant qubits, a SWAP gate is required. For example, to generate an interaction between $Q_1$ and $Q_3$, we first apply $U$ between $Q_2$ and $Q_3$ and then apply a SWAP gate between $Q_2$ and $Q_1$. Likewise with $Q_5$ and $Q_3$. The SWAP gate is realized by combining single-qubit gates and CZ gates as $SWAP=(I\otimes-Y/2)CZ(-Y/2\otimes Y/2)CZ(Y/2\otimes -Y/2)CZ(I\otimes Y/2)$ \cite{Barends2014}, where $Y/2$ ($-Y/2$) is $R_y(\pi/2)$ ($R_y(-\pi/2)$), representing the rotation by an angle $\pi/2$ ($-\pi/2$) about the $y$ axis.

The total depth of the sequences for the implementation of case 1a and 2 in the main text is 39, including 12 double-qubit-gate layers and 27 single-qubit-gate layers. For case 1b and case 3 in the main text, the total depths are 47, both including 12 double-qubit-gate layers and 35 single-qubit-gate layers. An example of the gate sequences for case 3 is shown in Fig. \ref{S1}.


\textbf{Gate optimization}

Calibrations and optimizations is a necessary step to successfully realize the theoretical circuits. Cross-talk on the Z control line \cite{Yan2019} is a source of error that needs to be firstly addressed. When CZ gate is applied, because of the $1\%-2\%$ Z cross-talk, it induces a frequency shift on other qubits, and leads unwanted dynamical phases. We correct these phase shifts by adding corresponding phase gates to each of them. Meanwhile, CZ gates must be applied in while all other qubits are idling. Secondly, due to the finite bandwidth and imperfection of the impedance matching in the route from the DAC channels to the qubit control lines, there is a pulse distortion after an applied pulse \cite{Neill2018,Kelly2014,Johnson2011d,Yan2019}. We use the deconvolution method to correct this kind of pulse distortion\cite{Johnson2011d,Yan2019}. Last, to mitigate the effects of dephasing, which produce more errors than energy relaxation in our experiment, we apply Hahn spin echoes \cite{PhysRev.80.580, Barends2014,Gustavsson2012,Martinis2003,Vion2003} to idling elements of the circuit.

\textbf{Numerical simulation with decoherence}

In order to evaluate the impact of decoherence, we use operator-sum representations to simulate the evolution of the system with the relaxation and the dephasing. In this method, we replace ideal matrices of quantum gates by Kraus operators with decoherence. Thus, the evolution of the system to which the quantum gate G is applied can be written as\cite{Nielsen2010}:
\begin{equation}
\varepsilon(\rho) = \sum_{k}E_{k} G\rho G^{\dagger} E^{\dagger}_{k}
\end{equation}
where G is the ideal matrix of the quantum gate, and $E_{k}$ is the operation elements for the decoherence. In this way, we can get the theoretical end states with decoherence after the designed circuits.

As for the specific matrix form of the operation elements, the operation elements are different for different kinds of decoherence. For the relaxation, the operation elements are 
\begin{equation}
\centering 
E_{1}=\left[
\begin{array}{cc}  
1 & 0 \\ 
0 & \sqrt{1-\gamma} 
\end{array}
\right ]\quad\quad
E_{2}=\left[
\begin{array}{cc}  
0 & \sqrt{\gamma} \\ 
0 & 0 
\end{array}
\right ]		
\end{equation}
where $\gamma$ is the probability of losing a exciton and is equals to the ratio of the gate time to the relaxation time $T_{1}$ in simulation. In addition, the operation elements for the dephasing are
\begin{equation}
\centering 
E_{3}=\left[
\begin{array}{cc}  
1 & 0 \\ 
0 & \sqrt{1-\gamma_{\phi}} 
\end{array}
\right ]\quad\quad
E_{4}=\left[
\begin{array}{cc}  
0 & 0 \\ 
0 & \sqrt{\gamma_{\phi} }
\end{array}
\right ]		
\end{equation}
where $\gamma_{\phi}$ is the probability that the exciton has been scattered and is equals to the ratio of the gate time to the dephasing time $T_{\phi}$ in simulation. Considering both relaxation and dephasing at the same time, we combine these operation elements like $E'_{1}=E_{1}E_{3}, E'_{2}=E_{1}E_{4}, E'_{3}=E_{2}E_{3}, E'_{4}=E_{2}E_{4}$. With these specific matrix form of the operation elements and operator-sum representations, we can simulate the evolution of the system and numerically get the states with decoherence.

\textbf{Quantum process tomography and bootstrapping}

In our experiment, the quantum process tomography (QPT) is used to identify the resetting process. In particular, we prepared the target qubit in different initial states: \ket{0}, \ket{1}, \ket{+}/\ket{-}, and \ket{+i}/\ket{-i}. At the end of the resetting protocol, we performed the quantum state tomography (QST) of the five-qubit system. The physical density matrices are then constructed via completely positive (CP) projection. After that, the successfully reset state is obtained by projecting the five-qubit density matrices into the success subspace $\mathcal{S}$. The density matrices of the target qubit after successful projection are then obtained by tracing out the probe qubits. With the density matrices started with four different initial states, the superoperator $\chi$ matrix is contracted. A completely positive and trace-preserving projection \cite{Knee2018} is used to ensure a physical estimation of the $\chi$ matrix.

We use bootstrapping technique to estimate the error bar in QPT. Based on the experimentally obtained five-qubit density matrices, we numerically sampled 200 sets of QST data with each initial states. In sampling the QST data, for each components of the measurement operators, we sampled 10,000 times and obtained the averaged population histograms. We used the sampled QST data to construct 200 sets of $\chi$ matrices, and the error bar is determined as the 1.96 times of the standard deviation of the process fidelities.

\section{Data Availability}
The experimental data for this experiment is available on Figshare, see \url{https://doi.org/10.6084/m9.figshare.12715493}.

\section{Code Availability}
The code for numerical simulations and analyzing the data of this experiment is also available on Figshare, see \url{https://doi.org/10.6084/m9.figshare.12715493}.

\section{ACKNOWLEDGEMENTS}
The authors thank the USTC Center for Micro- and Nanoscale Research and Fabrication, Institute of Physics CAS and National Center for Nanoscience and Technology for supporting the sample fabrication. The authors also thank QuantumCTek Co., Ltd. for supporting the fabrication and the maintenance of room temperature electronics. This research was supported by the National Key  R\&D Program of China (Grants No. 2017YFA0304300, No. 2018YFA0306703), the Chinese Academy of Sciences, and Shanghai Municipal Science and Technology Major Project (Grant No.2019SHZDZX01). This research was also supported by NSFC (Grants No. 11574380, No. 11905217) and Anhui Initiative in Quantum Information Technologies.

\section{AUTHOR CONTRIBUTIONS}
X.-B. Zhu, Y.-A. Chen and J.-W. Pan conceived the research. M. Gong, F.-H. Xu, Z.-D. Li, Z.-Z. Wang and X.-B. Zhu designed the experiment. S.-Y. Wang designed the sample. H. Deng, Z.-G. Yan and H. Rong prepared the sample. M. Gong, Y.-L. Wu, and S.-W. Li carried out the measurements. Y.-L. Wu developed the programming platform for measurements. Z.-D. Li, M. Gong, C. Zha, Y.-W. Zhao, and Y.-Z. Zhang did numerical simulations. M. Gong, Z.-D. Li, Z.-Z. Wang and C. Zha analyzed the results. F.-T. Liang, J. Lin, Y. Xu, C. Guo, L.-H. Sun and C.-Z. Peng developed room temperature electronics equipments. All authors contributed to discussions of the results and development of manuscript. X.-B. Zhu and J.-W. Pan supervised the whole project. 

\section{Competing interests}
The authors declare that there are no competing interests.

\nolinenumbers

\end{document}


\title{{\Large Supplementary Information for} \\ {\large \red{\textbf{Verification of a resetting protocol for an uncontrolled superconducting qubit}}}}	
	\author{Ming Gong,$^{1,2,3}$ Feihu Xu,$^{1,2,3}$ Zheng-Da Li,$^{1,2,3}$ Zizhu Wang,$^{4}$ Yu-Zhe Zhang,$^{1,2,3}$ Yulin Wu,$^{1,2,3}$ Shaowei Li,$^{1,2,3}$  Youwei Zhao,$^{1,2,3}$ Shiyu Wang,$^{1,2,3}$ Chen Zha,$^{1,2,3}$ Hui Deng,$^{1,2,3}$ Zhiguang Yan,$^{1,2,3}$ Hao Rong,$^{1,2,3}$  Futian Liang,$^{1,2,3}$ Jin Lin,$^{1,2,3}$ Yu Xu,$^{1,2,3}$ Cheng Guo,$^{1,2,3}$ Lihua Sun,$^{1,2,3}$ Anthony D. Castellano,$^{1,2,3}$ Cheng-Zhi Peng,$^{1,2,3}$  Yu-Ao Chen,$^{1,2,3,}$ Xiaobo Zhu,$^{1,2,3,}$\footnote{Corresponding author. Email: xbzhu16@ustc.edu.cn} and Jian-Wei Pan$^{1,2,3}$\vspace{0.2cm}}
\affiliation{$^1$  Hefei National Laboratory for Physical Sciences at the Microscale and Department of Modern Physics, University of Science and Technology of China, Hefei 230026, China}
\affiliation{$^2$  Shanghai Branch, CAS Center for Excellence in Quantum Information and Quantum Physics, University of Science and Technology of China, Shanghai 201315, China}
\affiliation{$^3$  Shanghai Research Center for Quantum Sciences, Shanghai 201315, China}
\affiliation{$^4$  Institute of Fundamental and Frontier Sciences, University of Electronic Science and Technology of China, Chengdu 610054, China}

\maketitle

\beginsupplement

\section{Supplementary Note 1}

In Supplementary Table~\ref{details} we presented the detailed data in three subcases, i.e., case 1a, 1b, and 1c. Fore case 1a, 8 rotation angles are tested and all trace distances between the reset-state and the ideal original state are below 0.23. In case 1b, 4 rotation angles are tested, and the trace distances are below 0.34. We note that the difference in reset-state fidelity between case 1a and 1b mostly comes from the depth. The interaction unitary between case 1a and 1b are different, thus the final circuits to perform the tests have different depths. In case 1b, 8 more single-qubit-gate layers are applied, introducing more gate errors and decoherence effect. In case 1c, the target is initialized in $\ket{-}$ and decohered for $1\mu s$ before the other parts of state preparation. Before the application of free evolution, the trace distance of the target is measured to be around 0.10, indicating that the initial state has already been a mixed state. In this subcase, we measure 4 different rotation angles and obtained the trace distances being below 0.25.

\begin{table}[H]
	\resizebox{\textwidth}{!}{%
		\begin{tabular}{c|c|c|c|c|c|c|c}
			\hline\hline
			\begin{tabular}[c]{@{}c@{}}Initial\\ target state\end{tabular} & \begin{tabular}[c]{@{}c@{}}Initial state\\ trace distance\end{tabular} & \begin{tabular}[c]{@{}c@{}}Circuit\\ depth\end{tabular} & $U$ & $R$ & $\varphi /\pi$ & \begin{tabular}[c]{@{}c@{}}Reset state\\ trace distance\end{tabular} & \begin{tabular}[c]{@{}c@{}}Success\\ probability\end{tabular} \\ \hline
			\multirow{8}{*}{$|-\rangle$} & \multirow{8}{*}{0.026} & \multirow{8}{*}{\begin{tabular}[c]{@{}c@{}}27 (single-qubit)\\ 12 (double-qubit)\end{tabular}} & \multirow{8}{*}{$\frac{X\otimes Z+iY\otimes X}{\sqrt{2}}$} & \multirow{8}{*}{$R_z(\varphi)$} & 1/16 & 0.177 & 0.518 \\
			&  &  &  &  & 2/16 & 0.163 & 0.525 \\
			&  &  &  &  & 3/16 & 0.163 & 0.527 \\
			&  &  &  &  & 4/16 & 0.195 & 0.520 \\
			&  &  &  &  & 5/16 & 0.221 & 0.534 \\
			&  &  &  &  & 6/16 & 0.214 & 0.544 \\
			&  &  &  &  & 7/16 & 0.191 & 0.557 \\
			&  &  &  &  & 8/16 & 0.187 & 0.544 \\ \hline
			\multirow{4}{*}{$|1\rangle$} & \multirow{4}{*}{0.039} & \multirow{4}{*}{\begin{tabular}[c]{@{}c@{}}35 (single-qubit)\\ 12 (double-qubit)\end{tabular}} & \multirow{4}{*}{$\frac{-Z\otimes Z+iY\otimes X}{\sqrt{2}}$} & \multirow{4}{*}{$R_x(\varphi)$} & 5/16 & 0.284 & 0.455 \\
			&  &  &  &  & 6/16 & 0.277 & 0.467 \\
			&  &  &  &  & 7/16 & 0.302 & 0.442 \\
			&  &  &  &  & 8/16 & 0.333 & 0.451 \\ \hline
			\multirow{4}{*}{$\sum_i p_i |i\rangle \langle i| , \rho^2 < 1$} & \multirow{4}{*}{0.098} & \multirow{4}{*}{\begin{tabular}[c]{@{}c@{}}29 (single-qubit)\\ 12 (double-qubit)\end{tabular}} & \multirow{4}{*}{$\frac{X\otimes Z+iY\otimes X}{\sqrt{2}}$} & \multirow{4}{*}{$R_z(\varphi)$} & 5/16 & 0.272 & 0.453 \\
			&  &  &  &  & 6/16 & 0.246 & 0.458 \\
			&  &  &  &  & 7/16 & 0.228 & 0.449 \\
			&  &  &  &  & 8/16 & 0.245 & 0.451 \\ \hline\hline
		\end{tabular}%
	}
	\caption{Detailed data in case 1 in the main text. Three parts from top to bottom correspond to case 1a, 1b, and 1c, respectively. }
	\label{details}
\end{table}